\documentclass[reprint,
amsmath,amssymb,
]{revtex4-2}

\usepackage{graphicx}
\usepackage{dcolumn}
\usepackage{bm}
\usepackage{xcolor}

\begin{document}

\preprint{APS/123-QED}

\title{Dynamic modulation of dual-band nonreciprocal radiation in a graphene-Weyl semimetal plasmonic structure}

\author {Ye Ming Qing$^{1,2}$} 
\altaffiliation[Corresponding author: ]{ymqing@njupt.edu.cn}

\author{Jiao Liu$^{1}$}
\altaffiliation{Y. M. Qing and J. Liu contribute equally to this work.}

\author{Zhaoyan Yang$^{1}$}

\author{Yue Gou$^{1}$}

\author{Liang Wei Wu$^{3}$}

\author{Jun Wu$^{4}$}
\altaffiliation[Corresponding author: ]{mailswj2011@163.com}

\affiliation{$^{1}$College of Electronic and Optical Engineering $\&$ College of Flexible Electronics (Future Technology), Nanjing University of Posts and Telecommunications, Nanjing 210023, China}
\affiliation{$^{2}$State Key Laboratory of Millimeter Waves, Southeast University, Nanjing 210096, China}
\affiliation{$^{3}$School of Microelectronics, Hefei University of Technology, Hefei 230601, China}
\affiliation{$^{4}$College of Electrical Engineering, Anhui Polytechnic University, Wuhu 241000, China}

\date{\today}

\begin{abstract}
We introduce and develop a hybrid structure combining graphene and Weyl semimetal, capable of achieving dynamically adjustable dual-band nonreciprocal radiation. The results reveal that the nonreciprocal radiation can be attributed to the synergistic interaction between resonance mode excitation and the unique properties of Weyl materials, with the electric field distribution providing further insights into the graphene plasmon modes involved. By exploiting the resonant characteristics of graphene plasmons, we demonstrate that strong nonreciprocal radiation can be effectively regulated through adjusting the grating's geometric parameters, while maintaining robustness over a wide range. Notably, substantial dynamic tuning of the resonant wavelength for nonreciprocal radiation is achievable by modulating the Fermi level of graphene. Our research results offer promising prospects for the developing of complex energy harvesting and conversion systems within advanced thermal frameworks.
\end{abstract}

\maketitle


\section{\label{sec:level1}INTRODUCTION}
Kirchhoff's law stands as a fundamental principle in the field of thermal radiation, asserting that the spectral directional emissivity $e$($\lambda$, $\theta$) of an object is precisely equal to its spectral directional absorptivity $\alpha$($\lambda$, $\theta$). This law underpins both theoretical and practical advancements in thermal radiation, including solar energy harvesting and photothermal conversion \cite{WU2021100388,doi:10.1021/acs.chemrev.3c00159}, radiative cooling \cite{ZHAO2019489,Zhu:14}, and thermal camouflage \cite{li2018structured}, which implicitly rely on its validity. However, recent scholarly endeavors have rigorously demonstrated that the establishment of Kirchhoff's law in thermal radiation is not an inevitable consequence of thermodynamic principles but rather a consequence of Lorentz reciprocity. In the absence of Lorentz reciprocity, Kirchhoff's law ceases to hold, leading to a disparity between the emissivity and absorptivity of an object at specific wavelengths and directions. This revelation challenges traditional understandings of radiative transfer processes and propels the development of novel theories and applications. Nonreciprocal thermal radiation offers the potential for independent modulation of emissivity and absorptivity, thereby overcoming inherent energy losses associated with traditional equality between absorption and emission, and promising to push energy harvesting or conversion efficiencies closer to thermodynamic limits \cite{yang2024nonreciprocal,PhysRevApplied.18.027001,PhysRevApplied.18.034083,PhysRevX.12.021023,10.1063/5.0225127,WU2024125923}.

In recent years, researchers have explored and realized nonreciprocal radiation through various approaches, with InAs-based magneto-optical (MO) materials garnering significant attention \cite{wu2023enhancement,doi:10.1126/sciadv.abm4308,WU2023106794,FANG2024108602,esee8c442,LI2024107772}. In 2014, Zhu et al. pioneeringly demonstrated the excitation of strongly localized asymmetric guided modes using n-InAs magnetic photonic crystals, proving the feasibility of violating Kirchhoff's law of thermal radiation, albeit requiring an external magnetic field of 3 Tesla (T) \cite{PhysRevB.90.220301}. Subsequently, Zhao et al. numerically demonstrated nonreciprocal radiation by sandwiching thick InAs films between a dielectric grating and a continuous metallic base, thereby minimizing the necessary external magnetic field to 0.3 T \cite{Zhao:19}. Shayegan et al. experimentally validated the potential of MO photonic structures for realizing nonreciprocal radiation \cite{shayegan2023direct}. Liu et al. further validated the concept of broadband non-reciprocal absorption by utilizing magnetized gradient structures with epsilon-near-zero films \cite{liu2023broadband}. However, these structures inevitably necessitate the introduction of external magnetic fields, significantly complicating practical applications.

Recently, Weyl semimetals (WS)-based nonreciprocal devices have attracted considerable interest due to their internally existing Weyl points with opposite chirality. These Weyl points induce an anomalous Hall effect, resulting in an asymmetric dielectric tensor and thus exhibiting optical nonreciprocity without the need for external magnetic fields, greatly simplifying application conditions \cite{guo2023light,10.1063/5.0180575,Wang:24}. Zhao et al. confirmed that magnetic WS can be utilized to construct nonreciprocal thermal radiation devices \cite{doi:10.1021/acs.nanolett.9b05179}. Asadchy et al. designed small-sized nonreciprocal optical isolators using WS \cite{https://doi.org/10.1002/adom.202000100}. Furthermore, several researchers have integrated WS into photonic crystal structures to achieve multichannel nonreciprocal radiation \cite{10.1063/5.0109402,WU2023103161,LUO2023124259}. Wu and Qing utilized a cascaded grating structure incorporating WS to realize broadband and wide-angle nonreciprocal radiation \cite{10.1063/5.0134234}. However, these nonreciprocal devices often lack tunability and flexibility. To overcome this limitation, in our previous work \cite{QING2024107657,WU2024107187,D2CP05945B}, we integrated tunable material (i.e., graphene) into WS-based structures, achieving a certain degree of tunability, but the modulation wavelength range was limited and needed to be expanded.

In this study, we introduce and meticulously design a hybrid graphene-Weyl structure that offers unprecedented capabilities for dynamically tunable dual-band nonreciprocal radiation. Our findings underscore that the emergence of nonreciprocal radiation stems from the synergistic interplay between resonance mode excitation and the distinctive attributes of Weyl materials, with electric field distributions offering deeper insights into the participating graphene plasmon modes. By harnessing the resonant properties of graphene plasmons, we exhibit that robust and effective regulation of strong nonreciprocal radiation can be achieved through strategic adjustments to the grating's geometric parameters. Furthermore, we demonstrate remarkable dynamic tuning of the resonant wavelength for nonreciprocal radiation by modulating the Fermi level of graphene. These findings possess the potential to revolutionize the progress of sophisticated energy harvesting and conversion frameworks within advanced thermal systems, thereby enabling innovative applications and enhanced functionalities.

\section{Design and Theory}

\begin{figure}[b]
\centering
\includegraphics[width=8.5cm]{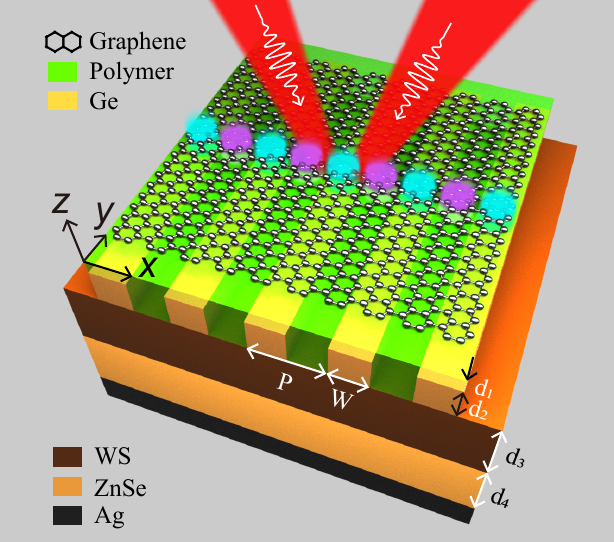}
\caption{Schematic diagram of the nonreciprocal graphene-WS plasmonic structure.}
\end{figure}

Figure 1 shows the designed structure, which consists of a silver substrate, a dielectric spacer, a WS layer, a grating layer, and a graphene monolayer from bottom to top, with a polymer dielectric serving as a buffer layer between the graphene and the grating layer. The specific structural parameters are set as follows: $P$ = 0.58 $\mu$m, $W$ = 0.26 $\mu$m, $d_{1}$ = 12 nm, $d_{2}$ = 0.12 $\mu$m, $d_{3}$ = 0.26 $\mu$m, and $d_{4}$ = 1.75 $\mu$m. The permittivity of silver ($\varepsilon_{Ag}$) is calculated from the Drude model \cite{ZHAO201481}: $\varepsilon_{Ag}=\varepsilon_\infty- \omega_p^2 / \left(\omega^2+j\omega\Gamma\right)$, where $\varepsilon_\infty=3.4$, $\Gamma=2.7\times{10}^{13}$ rad/s and $\omega_p=1.39\times{10}^{16}$ rad/s. The dielectric spacer is ZnSe with a refractive index of 2.4. The grating layer and the buffer layer are assumed to be Ge and polymer (i.e., NFC, a frequently utilized derivative of polyhydroxystyrene), with permittivity values of 16 and 2.4, respectively. Within the infrared frequency spectrum, the surface conductivity ($\sigma_g$) of monolayer graphene can be modeled using a Drude-like formula \cite{doi:10.1021/acsphotonics.9b00956}: $\varepsilon_g=1+i{\sigma_g}/{\left(\varepsilon_0\omega t_g\right)}$. Here, the equivalent permittivity ($\varepsilon_g$) of graphene is expressed in terms of the angular frequency ($\omega$), the permittivity of free space ($\varepsilon_0$), and the thickness of the graphene monolayer ($t_g$ = 0.34 nm). Notably, the conductivity of this material is significantly influenced by the Fermi energy ($E_f$), which we have defaulted to 1.2 eV in our study. The Fermi energy can be adjusted by applying a gate voltage or through the use of chemical doping methods. All other graphene-related parameters used here adhere to those detailed in our prior research, unless specified otherwise. The dielectric constant tensor for WS is characterized as \cite{doi:10.1021/acs.nanolett.9b05179}:
\begin{eqnarray}
\varepsilon = \begin{bmatrix}
\varepsilon _{d} & 0 & i\varepsilon _{a} \\
0 & \varepsilon _{d} & 0 \\
-i\varepsilon _{a} & 0 & \varepsilon _{d} \\
\end{bmatrix}
\end{eqnarray}
where the off-diagonal terms $\varepsilon_a={be^2}/{2\pi^2\hbar\omega}$. These non-zero, asymmetric terms violate the Lorentz reciprocity theorem, thereby potentially disrupting the conventional Kirchhoff's laws. The diagonal terms can be described by the Kubo-Greenwood formalism:
\begin{eqnarray}
\varepsilon_d=\varepsilon_b+i\frac{\sigma}{\varepsilon_0\omega}
\end{eqnarray}
where $\varepsilon_b$ is the background permittivity and $\sigma$ is the bulk conductivity given by:
\begin{eqnarray}
\sigma = && \frac{r_{s}g}{6}\Omega G\left ( \frac{\Omega }{2} \right )+i\frac{r_{s}g}{6\pi }\Bigg\{\frac{4}{\Omega } \left [ 1+\frac{\pi ^{2}}{3}\left ( \frac{k_{B}T}{E_{F}(T)} \right )^{2} \right ] \nonumber\\
&& + 8\Omega \int_{0}^{\xi _{c}}\frac{G(\xi)-G(\frac{\Omega }{2})}{\Omega ^{2}-4\xi ^{2}}\xi d\xi  \Bigg\}
\end{eqnarray}
where $r_s$, $\mathrm{\Omega}$, and $g$ represent the effective fine structure constant, the normalized frequency in complex form, and the number of Weyl nodes, respectively. $G(E)=n(-E)-n(E)$, where $n(E)$ is the Fermi distribution, $E_F(T)$ is the chemical potential. $\xi_c={E_c}/{E_F}$ is the normalized cutoff energy. The specific values of the parameters utilized in this study are available in prior research \cite{doi:10.1021/acs.nanolett.9b05179}. Here we only consider obliquely incident TM polarized light with an incident angle of $\theta$. Since the metal mirror at the bottom is sufficiently thick, transmission does not need to be considered. Therefore, the absorptivity and emissivity in the spectral direction can be simplified as:
\begin{eqnarray}
\eta\left(\theta\right)=\alpha\left(\theta\right)-e\left(\theta\right)
\end{eqnarray}
where $\alpha\left(\theta\right)=1-R\left(\theta\right)$, $e\left(\theta\right)=1-R\left(-\theta\right)$, and $R\left(\theta\right)$ are the absorptivity, emissivity and reflectivity, respectively. In this work, We utilized the rigorous coupled wave analysis (RCWA) method to assess the electromagnetic response of the system.

\section{Results and Discussion}

Figure 2a illustrates the absorption and emission spectra of the designed structure, with and without graphene incorporation. Notably, in the presence of graphene, resonant effects occur within the structure, giving rise to two prominent resonant peaks at 10.62 $\mu$m (with an amplitude of 0.98) and 10.84 $\mu$m (with an amplitude of 0.94), respectively. The distinct difference between the absorption and emission peaks naturally leads to pronounced dual-band nonreciprocal peaks, with amplitudes of 0.89 and 0.83, as depicted in Figure 2b. Conversely, in the absence of graphene, no significant resonant absorption effect is observed, which consequently prevents the excitation of nonreciprocal radiation phenomena even with reciprocal materials. This indirectly underscores the significance of resonant effects in generating nonreciprocal phenomena. The insets in Figure 2b respectively showcase the electric field distributions corresponding to these two resonant positions. The incident light resonates with the structure and localizes near the graphene, exhibiting a 2$\pi$ phase shift within one period, indicative of the mode characteristics of graphene plasmons.

\begin{figure}[h]
\centering
\includegraphics[width=8.5cm]{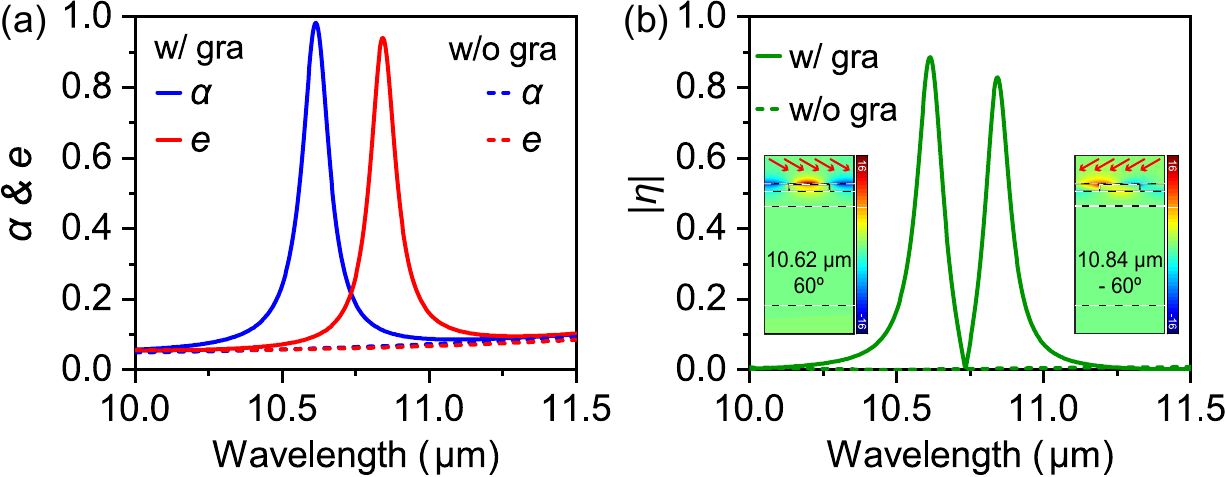}
\caption{(a) Absorptivity and emissivity as functions of wavelength at $\theta$ = 60$^\circ$ with and without graphene. (b) The corresponding nonreciprocal spectra. The insets show the electric field patterns ($E_x$) of the two resonance peaks.}
\end{figure}

We investigate the impact of several key structural parameters on the nonreciprocal performance, as illustrated in Figure 3. According to theoretical insights, the resonant wavelength of the graphene plasmons can be derived using the formula \cite{doi:10.1021/nn301888e,C4RA03431G}:
\begin{eqnarray}
\lambda=\frac{\sqrt2\pi\hbar c}{e}\sqrt{\frac{\beta\varepsilon_0\left(\varepsilon_{r1}+\varepsilon_{r2}\right)P}{E_f}}
\end{eqnarray}
where $\varepsilon_{r1}$ and $\varepsilon_{r2}$ represent the dielectric constants above and below the graphene, respectively, and $\beta$ is a dimensionless coefficient related to the structure. It is observed that as the $P$ increases, the resonant wavelength also increases, aligning with the characteristics of graphene plasmon modes, as shown in Figure 3a. Similarly, when the $W$ is increased, it effectively enhances the dielectric constant of the material beneath the graphene. Consequently, the resonant wavelength of the graphene plasmons also increases with $W$. As evident from Figures 3a and 3b, by adjusting the grating parameters, we can manipulate the graphene plasmons (as indicated by the star marks), thereby tuning the response wavelength of the nonreciprocal radiation. In contrast, modulating the two dielectric layers beneath the grating layer has a relatively minor impact on the nonreciprocal radiation, as demonstrated in Figures 3c and 3d. This observation further suggests that the mechanism underlying the nonreciprocal response in our designed structure originates from the excitation of graphene plasmons.

\begin{figure}
\centering
\includegraphics[width=8.5cm]{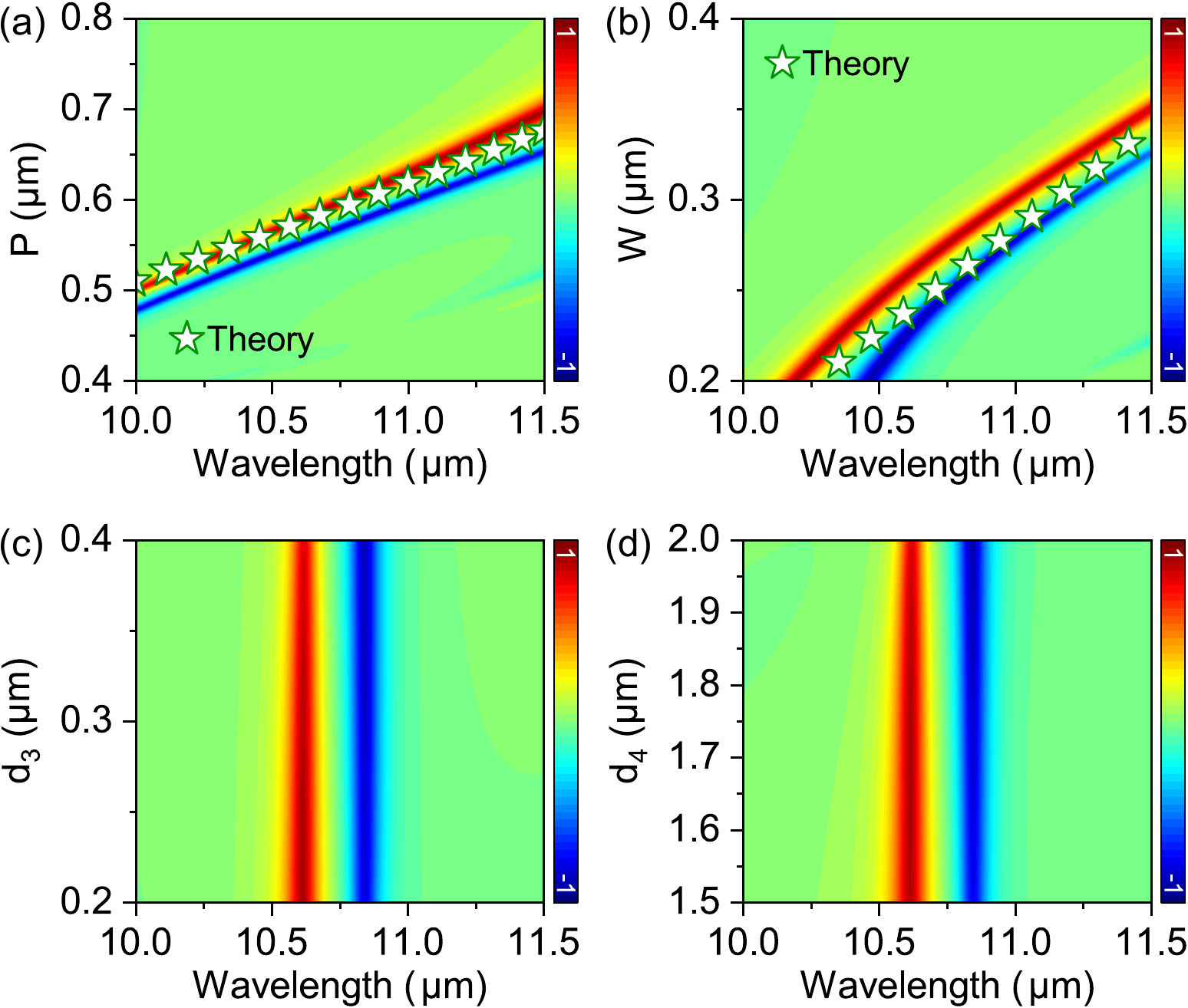}
\caption{Variations of the nonreciprocity for variation of geometric parameters: (a) grating period $P$; (b) graphene strip width $W$; and (c) WS thickness $d_3$, (d) ZnSe thickness $d_4$. The star markers are the predictions from theory. In each case, other parameters are not changed.}
\end{figure}

It is noteworthy that the incident angle is not explicitly represented in the formula 5, suggesting a minor influence of the incident angle on graphene plasmons. To further investigate the impact of the incident angle on the system's nonreciprocal properties, the results are illustrated in Figure 4. As expected, the incident angle exhibits a slight effect on the resonance position, with no significant shift in the resonance wavelength observed across varying incident angles. Moreover, within a broad range of incident angles, a nonreciprocal radiation intensity above 0.5 can still be maintained, as shown in Figure 4a. Specifically, we have analyzed the influence of the incident angle at two resonance wavelengths (i.e., 10.62 $\mu$m and 10.84 $\mu$m), as depicted in Figure 4b. Notably, within an incident angle range from nearly 40$^{\circ}$ to 80$^{\circ}$, a favorable nonreciprocal radiation intensity is sustained, demonstrating specific broadband nonreciprocal characteristics.

\begin{figure}[t]
\centering
\includegraphics[width=8.5cm]{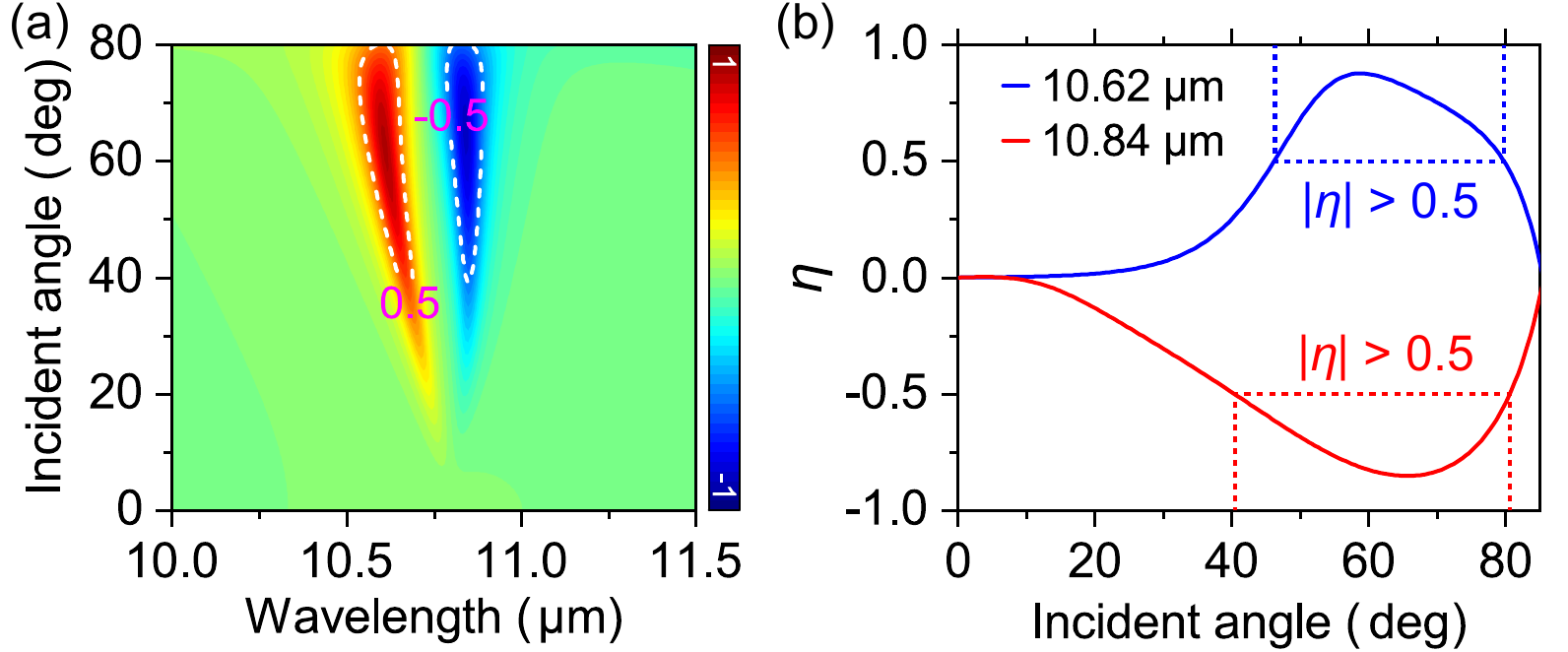}
\caption{(a) The calculated nonreciprocity vs the change of both wavelength and incident angle. (b) The spectral nonreciprocity vs the incident angle at the first (10.62 $\mu$m) and second resonances (10.84 $\mu$m).}
\end{figure}

\begin{figure}[t]
\centering
\includegraphics[width=8.5cm]{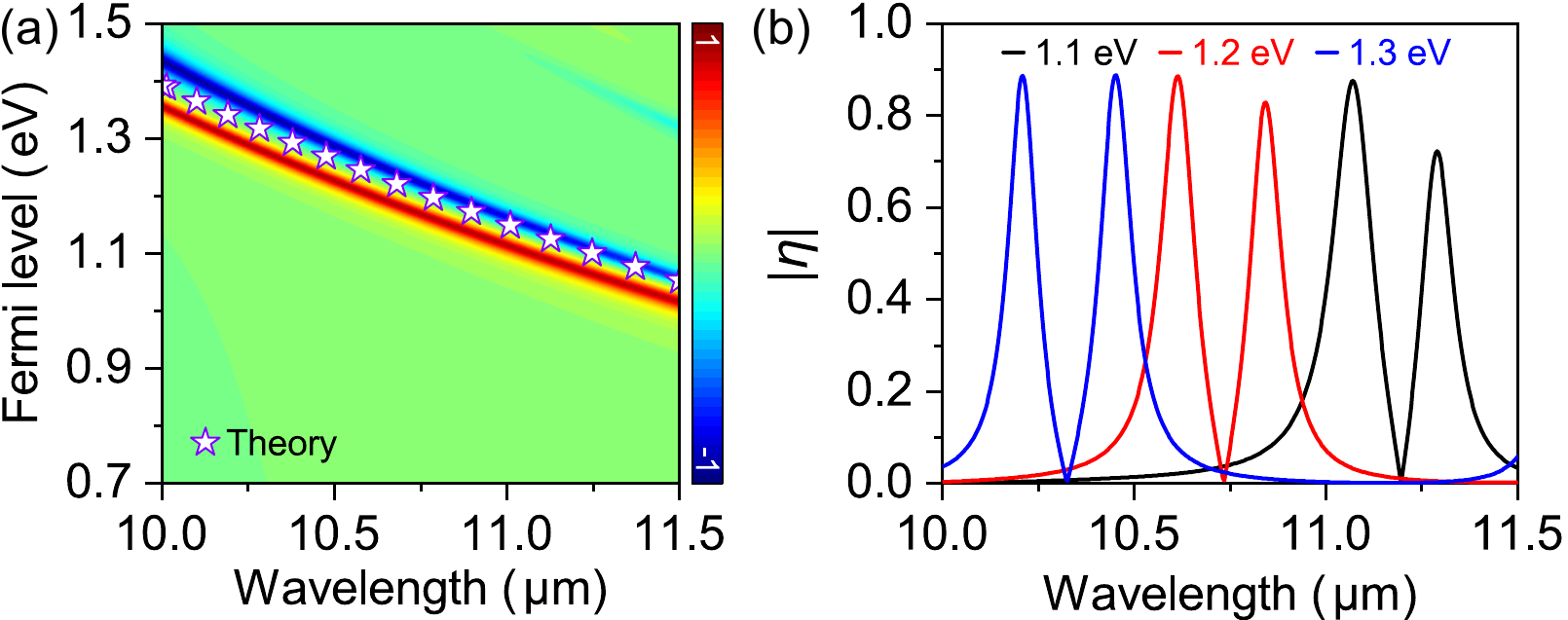}
\caption{(a) The influences of the Fermi level of graphene on nonreciprocity performance. The star markers are the predictions from theory. (b) The nonreciprocity spectra for different Fermi levels.}
\end{figure}

By adjusting the incident angle, we can dynamically control the intensity of nonreciprocal effects, yet we face limitations in modulating the resonant wavelength response. Fortunately, graphene plasmons offer a dynamic modulation pathway through tuning the Fermi level of graphene. As the Fermi level increases, the resonant wavelength decreases, as indicated by the star marks in Figure 5a. To elucidate this characteristic, we investigated the impact of the Fermi level on nonreciprocal properties. As anticipated, a significant blueshift in the resonant wavelength occurs with an increase in the Fermi level, while the intensity of nonreciprocal radiation remains relatively unchanged. This suggests that the resonant wavelength response can indeed be regulated by the Fermi level of graphene. We calculated the corresponding nonreciprocal radiation spectra for the system under three different Fermi levels. Compared to previous work, this study leverages the tunable properties of graphene plasmons effectively, achieving a modulation of nearly 400 nm in the resonant wavelength with a mere 0.1 eV change in the Fermi level. For instance, when the Fermi level is increased from 1.1 eV to 1.2 eV, the response wavelength of the left peak shifts from 10.21 $\mu$m to 10.62 $\mu$m, as shown in Figure 5b, demonstrating remarkable dynamic control capabilities.

\bigskip

\section{CONCLUSION}
In conclusion, we have successfully proposed and designed a hybrid graphene-WS structure that realizes dynamically tunable dual-band nonreciprocal radiation. Our findings have shown that the emergence of nonreciprocal radiation within this structure is attributed to the synergistic interplay between resonance mode excitation and the unique properties of WS. By harnessing the resonant characteristics of graphene plasmons, we have demonstrated that strong nonreciprocal radiation can be effectively regulated through strategic adjustments to the grating's geometric parameters, while maintaining robustness over a wide range. Furthermore, we have shown remarkable dynamic tuning of the resonant wavelength for nonreciprocal radiation by modulating the Fermi level of graphene, achieving a modulation of nearly 400 nm in the resonant wavelength with a mere 0.1 eV change in the Fermi level. Our results have provided deeper insights into the graphene plasmon modes involved in the process, with electric field distributions offering further evidence of the mode characteristics. The dual-band nonreciprocal radiation observed in our structure has significant potential to advance the design of sophisticated energy harvesting and conversion architectures within advanced thermal systems. By overcoming the limitations of previous nonreciprocal devices that lacked tunability and flexibility, our hybrid graphene-WS structure paves the way for innovative applications and enhanced performance in the field of thermal radiation.

\begin{acknowledgments}
The authors acknowledge the support of the National Natural Science Foundation of China (62305173, 62405142, 62401191), the Natural Science Foundation of the Jiangsu Higher Education Institutions of China (23KJB140015), the Open Project of the State Key Laboratory of Millimeter Waves (K202433), the Youth Talent Support Program of the Jiangsu Association for Science and Technology (JSTJ-2024-390), the Natural Science Research Start-up Foundation of Recruiting Talents of Nanjing University of Posts and Telecommunications (NY222080), and the Anhui Provincial Natural Science Foundation (Grant No. 2108085MF231).
\end{acknowledgments}

\nocite{*}

\bibliography{apssamp}

\end{document}